\documentclass[10pt,twocolumn,prl,aps,english,superscriptaddress]{revtex4-1}
\usepackage[T1]{fontenc}
\usepackage[utf8]{inputenc}
\usepackage{amsmath}
\usepackage{amssymb}
\usepackage{cancel}
\usepackage{graphicx}
\usepackage{esint}
\usepackage{multirow}
\usepackage[table,xcdraw]{xcolor}
\usepackage{makecell}
\usepackage{ulem}
\makeatletter

\@ifundefined{textcolor}{}{%
 \definecolor{BLACK}{gray}{0}
 \definecolor{WHITE}{gray}{1}
 \definecolor{black}{rgb}{1,0,0}
 \definecolor{GREEN}{rgb}{0,1,0}
 \definecolor{BLUE}{rgb}{0,0,1}
 \definecolor{CYAN}{cmyk}{1,0,0,0}
 \definecolor{MAGENTA}{cmyk}{0,1,0,0}
 \definecolor{YELLOW}{cmyk}{0,0,1,0}
}

\usepackage{dcolumn}
\usepackage{bm}
\usepackage{enumerate}

   \newcommand{\Moire}{Moir\'e }  
\newcommand{\moire}{Moir\'e }

\@ifundefined{showcaptionsetup}{}{%
\PassOptionsToPackage{caption=false}{subfig}}

\usepackage{babel}
\addto\captionsenglish{}
\renewcommand{\thefigure}{\@arabic\c@figure}

\makeatother

\begin{document}

\title{Moir\'e heterostructures as a condensed matter quantum simulator}

\author{Dante M. Kennes}
\altaffiliation{These authors contributed equally.\\ Email: Dante.Kennes@rwth-aachen.de}

\affiliation{Institut f\"ur Theorie der Statistischen Physik, RWTH Aachen University and JARA-Fundamentals of Future Information Technology, 52056 Aachen, Germany}
\affiliation{Max  Planck  Institute  for  the  Structure  and  Dynamics  of  Matter and Center Free-Electron Laser Science, Luruper  Chaussee  149,  22761  Hamburg,  Germany}

\author{Martin Claassen}
\altaffiliation{These authors contributed equally.\\ Email: Dante.Kennes@rwth-aachen.de}
\affiliation{Center  for  Computational  Quantum  Physics,  Flatiron  Institute,  New  York,  NY  10010  USA}
\affiliation{Department of Physics and Astronomy, University of Pennsylvania, Philadelphia, PA 19104, USA}

\author{Lede Xian}
\altaffiliation{These authors contributed equally.\\ Email: Dante.Kennes@rwth-aachen.de}
\affiliation{Songshan Lake Materials Laboratory, 523808 Dongguan, Guangdong, China} 
\affiliation{Max  Planck  Institute  for  the  Structure  and  Dynamics  of  Matter and Center Free-Electron Laser Science, Luruper  Chaussee  149,  22761  Hamburg,  Germany}

\author{Antoine Georges}
\affiliation{Coll{\`e}ge de France, 11 place Marcelin Berthelot, 75005 Paris, France}

\affiliation{Center for Computational Quantum Physics, Flatiron Institute, 162 Fifth Avenue, New York, NY 10010, USA}

\affiliation{CPHT, CNRS, {\'E}cole Polytechnique, IP Paris, F-91128 Palaiseau, France}

\affiliation{DQMP, Universit{\'e} de Gen{\`e}ve, 24 quai Ernest Ansermet, CH-1211 Gen{\`e}ve, Suisse}

\author{Andrew J. Millis}
\affiliation{Center  for  Computational  Quantum  Physics,  Flatiron  Institute,  New  York,  NY  10010  USA}
\affiliation{Department  of  Physics,  Columbia  University,  New  York,  NY,  10027  USA}

\author{James Hone}
\affiliation{Department of Mechanical Engineering, Columbia University, New York, NY, USA}

\author{Cory R. Dean}
\affiliation{Department  of  Physics,  Columbia  University,  New  York,  NY,  10027  USA}

\author{D. N. Basov}
\affiliation{Department  of  Physics,  Columbia  University,  New  York,  NY,  10027  USA}

\author{Abhay Pasupathy}
\email[Email: ]{apn2108@columbia.edu}
\affiliation{Department  of  Physics,  Columbia  University,  New  York,  NY,  10027  USA}

\author{Angel Rubio}
\email[Email: ]{Angel.Rubio@mpsd.mpg.de}
\affiliation{Max  Planck  Institute  for  the  Structure  and  Dynamics  of  Matter and Center Free-Electron Laser Science, Luruper  Chaussee  149,  22761  Hamburg,  Germany}
\affiliation{Center  for  Computational  Quantum  Physics,  Flatiron  Institute,  New  York,  NY  10010  USA}
\affiliation{Nano-Bio  Spectroscopy  Group,  
Universidad  del  Pa\'is  Vasco,  20018  San  Sebastian,  Spain}

\begin{abstract}
{\bf 
Twisted van der Waals heterostructures have latterly received prominent attention for their many remarkable experimental properties, and the promise that they hold for realising elusive states of matter in the laboratory. We propose that these systems can, in fact, be used as a robust quantum simulation platform that enables the study of strongly correlated physics and topology in quantum materials. Among the features that make these materials a versatile toolbox are the tunability of their properties through readily accessible external parameters such as gating, straining, packing and twist angle; the feasibility to realize and control a large number of fundamental many-body quantum models relevant in the field of condensed-matter physics; and finally, the availability of experimental readout protocols that directly map their rich phase diagrams in and out of equilibrium. This general framework makes it possible to robustly realize and functionalize new phases of matter in a
modular fashion, thus broadening the landscape of accessible physics and holding promise for future technological applications.

}
\end{abstract}

\maketitle

\section*{Introduction}

The inseparability of groups of particles into products of their individual states underpins the most counter-intuitive predictions of quantum mechanics. In condensed matter physics, recent focus has been placed on finding and understanding quantum materials -- systems in which a delicate balance between the crystal lattice and strong electronic interactions lead to new inseparable, collective behaviors with exotic properties. Encompassing rich phenomenologies ranging from superconductivity at high temperatures to topologically-ordered states of matter with fractionalized excitations, numerous materials have recently come under intense scrutiny for the realization of quantum phases on demand \cite{Basov2017}.

A major impediment  to the systematic study of correlated electron physics is the comparative lack of tunability of conventional chemical compounds. The main means of control are  pressure, strain and doping. This limited range of control knobs  severely limits experimental exploration of the wider phase diagram, essential to guide the discovery of novel quantum phases and to controlled studies of quantum criticality, unconventional phase transitions, or realizations of exotic topological states of matter that are both of fundamental interest and could have applications in emerging quantum technologies such as quantum computing and simulation.

Another impediment lies in the difficulty of solving these paradigmatic quantum Hamiltonians via present computational approaches. As an alternative, in the early 80's Feynman proposed quantum simulation as a paradigm to turning the problem upside down \cite{feynman1982}: implementing clean models using real physical systems, to ``simulate'' the ground state, thermodynamic behavior or non-equilibrium dynamics of such models. If the implementation is well-controlled, the results can be used to disentangle the more complex material's phenomena for which the original model was set up, as well as to guide the way towards stabilizing and controlling new and exotic phases of matter in real quantum materials. 

{
Much progress has been achieved via studying ultracold gases of bosonic and fermionic atoms, which can be confined in optical lattices to realize lattice models of condensed matter physics in a controlled manner \cite{bloch2012quantum}. While these systems are highly controllable and can faithfully realize certain quantum Hamiltonians, it remains an experimental challenge to engineer Hamiltonians with tunable long-range interactions and especially to access low temperature emergent long-range ordered phases, such as e.g. the intriguing d-wave state of the t-t' repulsive Hubbard model \cite{Zheng1155}. 

This article identifies Moir\'e heterostructures of van der Waals materials as an alternative and complementary condensed matter approach to realize a large set of highly controllable quantum Hamiltonians. While such systems do not typically afford the high level of isolation and precise tunability that cold atomic systems do, the large degree of tunability control available through readily accessible experimental knobs allows for the exploration of phase diagrams of vast and novel sets of many-body Hamiltonians in and out of equilibrium and at very low temperatures. In particular, we will review how different choices of two-dimensional (2D) heterostructure compositions at different twist angles realize a wide range of effective low-energy electronic tight-binding Hamiltonians with various geometries, dimensionality, frustration, spin-orbit coupling as well as interactions  beyond the Hubbard limit, including partially screened and  unscreened Coulomb interactions. This allows for flexible interaction-engineering as well as control of band structure and topology.
}

\section*{Realizing Model Quantum Hamiltonians in van der Waals Heterostructures}

 Moir\'e heterostructures of van der Waals materials exploit quantum interference to quench the effective kinetic energy scales, which permits both driving the system at low energies to an interaction-dominated regime, and drastically enhancing anisotropies or reducing symmetries of the monolayer. Conceptually, the physical mechanism can be understood straightforwardly in analogy to classical Moir\'e interference patterns (see Fig. 1, central inset): if two identical patterns with a discrete translation symmetry in two dimensions are superimposed at an angle, the combined motif retains a periodicity at much longer wave lengths for an infinite set of small rational angles.

Similarly, a heterostructure of two or more monolayers with commensurate lattice constants stacked at a twist retains spatial periodicity for small commensurate angles, albeit with a much larger Moir\'e unit cell. This Moir\'e superlattice can span many unit cells and defines a new crystal structure with a mini Brillouin zone. Within this Moir\'e Brillouin zone, the folded bands lead to a multitude of crossings, which are subsequently split via interlayer hybridization \cite{bistritzer2011}. Crucially, lattice relaxation enhances these avoided crossings and typically segregates sets of almost-dispersionless bands on meV energy scales \cite{yoo2019atomic}, which can be addressed individually via gate-tuning the chemical potential.

A complementary view of the physics of a Moir\'e band follows from considering its real space description. A localized electron in a Moir\'e band can be viewed as occupying a virtual ``Moir\'e orbital'' that can extend over hundreds of atoms, effectively spreading over the entire Moir\'e unit cell. 
Importantly, the shape of Moir\'e orbitals changes with the size of the Moir\'e unit cell so that the electronic interactions become a function of the twist angle as well, decreasing at a slower rate than the kinetic energy scales for decreasing angle. This permits tuning \textit{a priori} weakly-interacting electronic systems into regimes dominated by electronic correlations in a controlled fashion. 


A first demonstration of the above phenomenology was realized in materials with hexagonal structure and multiple atomic orbitals. The Moir\'e band width for twisted bilayers of graphene does not behave monotonically as a function of twist angle but exhibits a series of ``magic'' angles \cite{bistritzer2011} at which the Moir\'e bands near charge neutrality become almost dispersionless over a large fraction of the Moir\'e Brillouin zone.

More broadly, as reductions of the electronic band width equivalently enhance the role of competing energy scales, this microscopic structural knob to selectively quench the kinetic energy scales in 2D materials opens up possibilities to selectively engineer heterostructures with properties that are dominated via many-body electronic interactions, ultra-strong spin-orbit interactions from heavier elements, electron-lattice interactions or electron-photon interactions, permitting realization of a wide range of novel correlated or topological phenomena. { Here, we  provide a perspective on how different  universal classes of quantum many-body Hamiltonians can be engineered and controlled using \moire heterostructures}

\section*{ The Moir\'e Quantum Simulator: Interacting Electrons in Tailored Lattices}

Table~I summarizes some of the realizable lattice structures, associated model Hamiltonians and featured quantum phases that can be achieved within a 2D twisted van der Waals heterostructure framework. In our discussion, we will concentrate first on those systems realized or proposed in the literature, summarized in the innermost circle of Fig.\ref{fig1}. Then, we will provide a perspective on future potential control and simulation possibilities, shown in the outer circle of Fig.\ref{fig1}.

\subsection*{Honeycomb Lattice -- Twisted Bilayer Graphene, MoS$_2$ and others}
So far, most experimental and theoretical research has concentrated on twisted bilayer graphene (TBG) 
(see \cite{cao2018a,cao2018b}, and the review \cite{CoryP} for an overview on TBG), with extensions to triple- and quadruple-layer generalizations \cite{chen2019signatures,chen2019evidence,liu19,shen19,cao2019electric,tutuc2019,he2020tunable}. 
TBG realizes a Moir\'e superstructure with the relevant low-energy degrees of freedom effectively again forming a honeycomb lattice, although at meV energy scales and imbued with an additional orbital degree of freedom reflecting the two layers of the original system\cite{Yuan18,Koshino18}. 

{
\begin{table*}[t!]
{\scriptsize
\begin{tabular}{|c|c|c|c|}
\hline
\multicolumn{4}{|l|}{\cellcolor[HTML]{c9f6ff} Twisted heterostructures of weakly-correlated van-der-Waals monolayers } \\ \hline
     \textbf{\textit{Lattice}}  &    \textbf{\textit{Model}}  &   \textbf{\textit{Possible materials realizations}} & \textbf{\textit{Correlated phases}}      \\ \hline
     
    honeycomb  &    \makecell{two-orbital extended Hubbard model \cite{Koshino18} \\ fragile topological insulator \cite{po18a} }  & \makecell{twisted bilayer graphene \\ (BN substrate, with/without twist)}    &   \makecell{Mott insulation \cite{cao2018b} \\ superconductivity \cite{cao2018a} \\ correlated QAH insulator \cite{sharpe19,serlin19} }   \\ \cline{2-4}
      &       two-orbital extended Hubbard model   &    twisted double bilayer graphene  &   \makecell{ferromagnetic insulator \\ superconductivity \cite{liu19,shen19} \\ triplet pairing \cite{lee19} }   \\ \cline{2-4}
      &    asymmetric  $p_x,p_y$ Hubbard model {\cite{xian20,2008.01735}} &    twisted bilayer MoS$_2$, MoSe$_2$  &   {nematic (anti)ferromagnets \cite{xian20}}    \\ \cline{2-4}      
      & domain wall networks & \makecell{small-angle twisted bilayer graphene \\ with domain reconstruction \cite{huang18b,efimkin18,kerelsky19,mcgilly19}} & \\ \hline
      
    triangular &    \makecell{Hubbard model \\ (with/without strong SOC)} & \makecell{twisted bilayer WS$_2$, WSe$_2$ \cite{wang19} \\ twisted WS$_2$/WSe$_2$ heterostructures \cite{regan2020,tang2020} \\ twisted double bilayers of WSe$_2$ \cite{an19}}  & \makecell{correlated insulator \cite{wang19} \\ superconductivity{?} \\ Wigner crystals \cite{regan2020}} \\ \cline{2-4}
    & doped multi-orbital Hubbard models & \makecell{twisted heterostructures of \\ MoS$_2$, WS$_2$, WSe$_2$} & \makecell{Moir\'e excitons {\cite{Jin2019,wang20,shimazaki2020strongly}}}
    \\ \cline{2-4}
    
     &    multi-orbital Kanamori models & twisted bilayer boron nitride & \makecell{spin density wave \\ $d$-wave superconductivity \cite{Xian18} }  \\ \hline

    rectangular &   \makecell{1D ionic Hubbard model \\ 1D--2D crossover} & twisted bilayer GeSe & \makecell{Luttinger liquid \\ Mott insulator \\ bond density waves \cite{kennes19} } \\ \cline{2-4}
      &   \makecell{inverted band insulator, strong SOC} & twisted bilayer WTe$_2$ & \makecell{ quantum spin Hall insulator, \\ fractional Chern/topological insulator} \\ \hline
    any &  Hofstaedter models & \makecell{twisted bilayer graphene or transition-metal \\ dichalcogenides in strong magnetic fields} & fractional Chern insulator \cite{spanton17} \\ \hline
    Kagome & Kagome Heisenberg model &  ?? &  \makecell{Z$_2$ QSL \\ U(1) QSL \\ quantum chiral spin liquid \\ valence bond crystal}   \\ \hline
    \makecell{decorated\\Kagome} & Hubbard model (putative??) & twisted bilayer MoS$_2$, MoSe$_2$ & ?? \\ \hline
    3D & flat-band Hubbard-Kanamori models & twisted multilayer ``staircase'' & ?? \\ \hline

\multicolumn{4}{|l|}{\cellcolor[HTML]{c9f6ff} Proximity Effects } \\ \hline       
     \textbf{\textit{Lattice}}  &  \textbf{\textit{Model}}  &   \textbf{\textit{Possible materials realizations}} & \textbf{\textit{Correlated phases}}      \\ \hline
     \makecell{honeycomb,\\triangular}  &  proximity-induced Rashba SOC     &   \makecell{twisted bilayer graphene on \\ WS$_2$, WSe$_2$ substrate {\cite{arora2020superconductivity}}}    &   correlated QSH insulator    \\ \hline
     \makecell{rectangular}  &  proximity-induced superconductivity   &  \makecell{superconductor, twisted bilayer GeSe, \\ TMDC ``sandwich'' heterostructure}   &   \makecell{1D Kitaev superconductor \\ Majorana bound states}     \\ \hline

\multicolumn{4}{|l|}{\cellcolor[HTML]{c9f6ff} Twisted heterostructures of correlated monolayers } \\ \hline       
    \textbf{\textit{Lattice}}  &  \textbf{\textit{Model}}  &   \textbf{\textit{Possible materials realizations}} & \textbf{\textit{Correlated phases}}      \\ \hline
    &  Moire ferromagnet {\cite{hejazi2020noncollinear}}  &  twisted odd-multilayer CrI$_3$ & \makecell{Moire domain wall \\ ferromagnets} \\ \hline
    &  Moire Kitaev model  &  twisted multilayer $\alpha$-RuCl$_3$ & \makecell{Kitaev QSL \\ stripe order \\ Majorana fermions} \\ \hline
    &  ?? &  twisted bilayer TaSe$_2$ & ?? \\ \hline
    &  ?? &  twisted bilayer NbSe$_2$ & ?? \\ \hline
\end{tabular}}
	\caption{Overview of possible quantum Hamiltonians, materials realizations and phases in twisted Moir\'e heterostructures.}
	\label{tab:over}
\end{table*}
}

\begin{figure*}[t!]
\includegraphics[width=0.95\linewidth,clip]{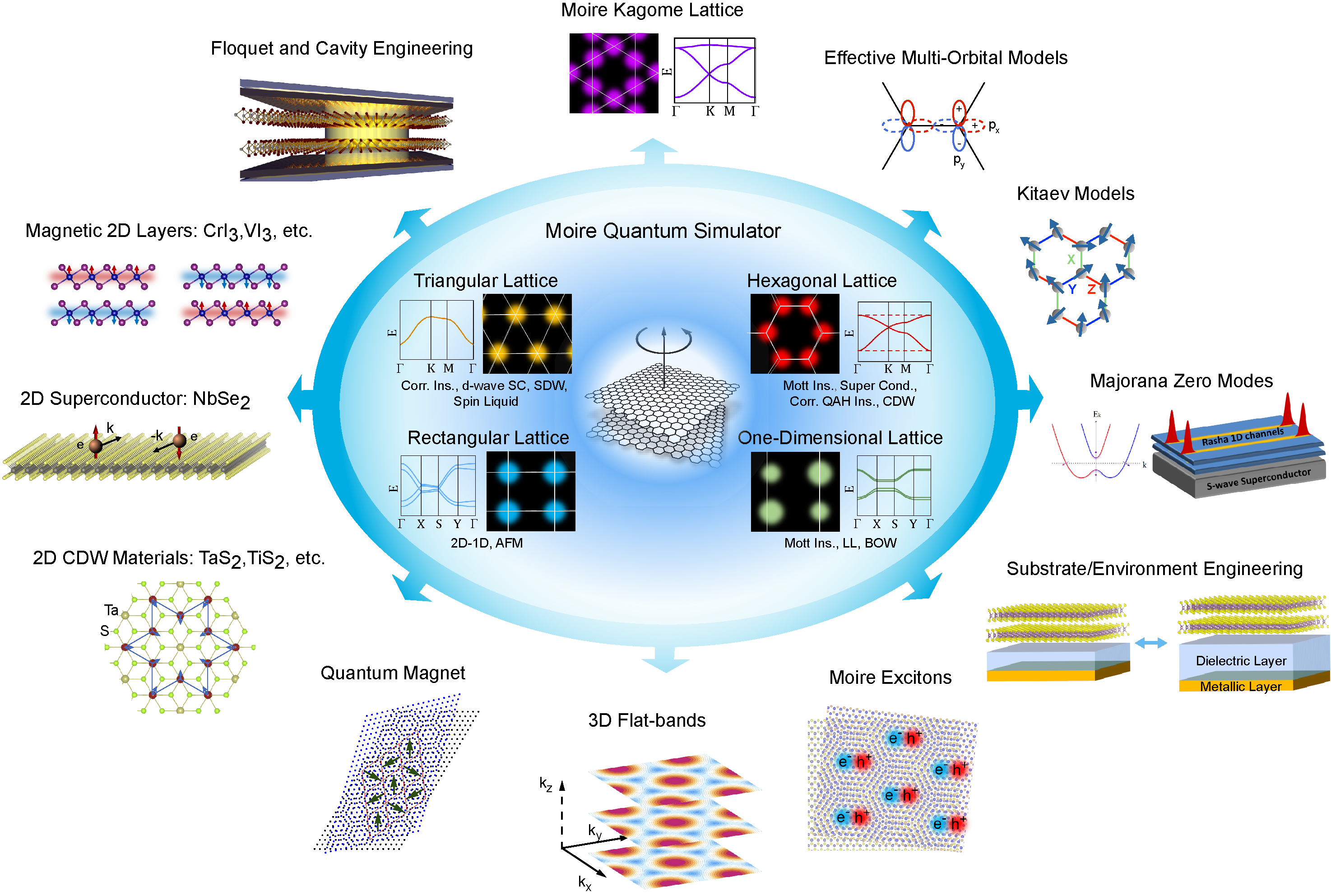}
\caption{{\bf Moir\'e Quantum Simulator:} Stacking sheets of van der Waals heterostructures with a twist gives rise to a plethora of effective low-energy Hamiltonians. Some of these realizations that were studied in the literature are given in the inner circle of the figure with the corresponding lattice, band structure and potential phases of matter that can be realized. For the red band structure the dashed lines indicate the flat bands additionally present in the case of twised bilayer MoS$_2$ compared to twisted bilayer graphene. In the outer circle we outline a perspective  on possible future directions of Moir\'e quantum simulators with many intriguing developments to be expected. }
\label{fig1}
\end{figure*}

With superconductivity \cite{cao2018a,Yankowitz18,lu2019superconductors,stepanov2019interplay}, correlated insulators \cite{cao2018b,lu2019superconductors,stepanov2019interplay} and the quantum anomalous Hall effect \cite{sharpe19,serlin19} already realized experimentally, an intriguing possibility indicated by recent theoretical analyses is that
repulsive interactions  favor topological $d+id$ instead of nodal $d$-wave pairing \cite{Xu18,Liu18,Kennes2018d0}. Such topological superconducting states are potentially relevant to topological quantum computing \cite{nayak07}, and can be harnessed and controlled via tailored laser pulses \cite{claassen18}. 


Expanding the catalog of engineered lattice structures, we next consider twisting two monolayers of MoS$_2$ \cite{Naik18,xian20,2008.01735}. In these structures families of flat bands emerge at the band edges, the first two of which realizes a single orbital Hubbard model while the second realizes a more exotic strongly asymmetric $p_x$-$p_y$ Hubbard model {\cite{xian20,2008.01735}} both on the Honeycomb lattice, see inner circle of Fig.~\ref{fig1}. The latter emerges at small twist angle, where the overall band width of these families of flat bands is tuned into the meV regime. Due to destructive interference within the strongly asymmetric $p_x$-$p_y$ Hubbard model itself ({ meaning that the two-orbitals have very different hopping amplitudes}), attached at the bottom and top of these flat bands one can find bands with even much lower dispersion. 
The asymmetric $p_x$-$p_y$ Hubbard model is therefore another lattice Hamiltonian that can effectively be engineered in this solid-state framework, and also controlled by external parameters, such as strain or fields. This could provide interesting insights into highly degenerate systems and the interplay of magnetism with these type of band structures. Similar phenomena are expected to emerge in other twisted TMDs homo-bilayers, such as twisted bilayer MoSe$_2$ and WS$_2$.

\subsection*{Triangular lattices -- BN, WSe$_2$}

A natural extension of TBG involves two twisted sheets of boron nitride (BN), which is often called white graphene for its structural similarity and large band gap. 
When twisting two BN sheets, the consequences are even more dramatic; while in graphene the quenching of the kinetic energy scales relies on being closed to a set of magic angles, in BN three families of flat band emerge, where the flattening of the bands is a monotonous function of twist angle. The effective lattice structure that is engineered  at low enough twist angle is effectively triangular \cite{Xian18,Ni2019}. 

A similar engineered lattice structure is obtained by twisting two sheets of tungsten diselenide WSe$_2$ \cite{wang19} or other transition metal dichalcogenide (TMD) heterostructures \cite{Wu17,regan2020,tang2020}. In all these cases a superstructure emerges which effectively confines the low energy physics onto a {\it triangular} lattice. In addition, the heavier transition metal elements in such TMD systems impose substantial spin-orbit coupling onto the low-energy degrees of freedom, { which can in turn induce interesting topological properties.} 

Fundamentally, triangular lattices are prototypical model systems to study the role of geometric frustration on electronic and magnetic orders. Using twisted materials one might thus simulate effectively the properties of these theoretically challenging models, relevant also for understanding many other condensed matter systems, such as BEDT-TTF organic compounds and their rich phase diagrams \cite{Shimizu03,Kagawa2005,Tajima06}. 
Debated correlated states of matter that can be realized this way range from 120$^\circ$ Neel ordered states to exotic, topological forms of superconductivity and quantum chiral spin liquids \cite{Szasz20}.

{
Indeed, in the case of TMD homobilayers, {features consistent with} superconducting phases {(with further experimental evidence for a truly superconducting state still highly desirable)} have already been seen in close proximity to the insulating phase \cite{wang19}. An advantage of these materials is that the angular precision needed in order for the emergence of new quantum phases is not as stringent as in graphene, so samples are easier to produce. However, 
the quality of the constituent materials tends to be poorer, and might need to be  improved to allow a well-controlled realization of the model Hamiltonians and phenomena described in Table~\ref{tab:over}  and Fig.~\ref{fig1}. 
}



\subsection*{Rectangular lattices --  2D to 1D crossover in GeSe or GeS}

While all of the examples above build on stacking monolayers with a three-fold (120$^\circ$) rotational symmetric lattice structure, this is certainly not the only interesting scenario that one can consider \cite{Kariyado19}. To illustrate this we next turn our attention to twisted bilayer germanium selenide (GeSe) \cite{kennes19} or similarly germanium sulfide (GeS). Monolayer GeSe has a rectangular unit cell with a 180$^\circ$ rotational symmetry. When two sheets of GeSe are twisted,  a similar quenching of the kinetic energy scales can be found as in the above discussed cases, however the effective lattice structure realized by such a system is a square lattice at small angles of 8$^\circ$-15$^\circ$  while at lower angles the hopping in one of the principal directions is reduced much more strongly than the corresponding hopping along the second principal direction. At about 6$^\circ$ the effective lattice system realized is entirely dispersion-less along one direction while it has a residual (though small due to the band narrowing at small angles) dispersion along the second one; see inner circle of Fig.~\ref{fig1}. Thus twist angle provides a control knob to tune the system from effectively two-dimensional to effectively one-dimensional low energy physics, providing new insights into this fundamental and poorly understood problem.

One-dimensional quantum systems exhibit strong collective effects. Bilayer GeSe at small twist angle should thus provide an experimental example to study this emergent quantum regime, where Luttinger liquid, Mott insulating and bond ordered wave states dominate the stage. Furthermore, sizable spin-orbit coupling of the low-energy bands emerges at larger twist angles, suggesting that interesting spin-momentum locked states with possibly nematic properties could be engineered \cite{Harter17,Harter18}. 



\section*{Experimental realizations and readout of the Moir\'e quantum simulator}

\subsection*{Experimental Realizations}


\Moire structures on layered materials have been observed by microscopy for at least the past three decades. In 3D crystals they can appear naturally during growth due to the presence of screw dislocations, or can be produced by mechanical damage of the crystal structure. These structures remained mostly curiosities until the advent of graphene, when it was realized that \moire structures are an effective way of tuning bandstructure (strong peaks in the local density of states were observed when twisting graphene on bulk graphite) \cite{jiang2019charge}. Such \moire patterns are commonly seen in chemical vapor deposition based grown multilayer graphene and TMD films, and many experiments to understand the properties of these patterns were performed in the past decade \cite{Li2010, Crommie2015} . In these samples, the twist angles generated were largely uncontrolled, and often showed disorder on sub-micron lengthscales, making the general applicability of these samples to other experiments limited.

The first steps on the road to making \moire heterostructures was established by mechanically stacking two monolayers \cite{Dean2010}  of  graphene and hBN to make high quality graphene layers for transport measurements. Due to the natural lattice mismatch between graphene and hBN, such structures naturally featured \moire textures \cite{Li2010}.
The most successful technique to realize twisted bilayer systems is the "tear and stack" method and its variants \cite{Tutuc2017} . This technique essentially separates a single monolayer into two pieces, which are then naturally aligned with each other. One of the pieces is picked up, rotated by the desired angle and stacked with the second piece to create the desired \moire structure. This simple and elegant technique has so far proved to be the most productive method, but has its limitations in reproducibility and uniformity of the \moire patterns produced (other techniques such as pushing with an atomic force microscopy tip \cite{Hone2019a} or introducing heteorostrain \cite{edelberg2019tunable} have been developed). The development of techniques to measure  the wavelength and uniformity of the \moire patterns in-situ are promising developments in the search to create the high-quality and precise \moire structures that we are discussing here. 

{
The in-plane bonding in two dimensional materials can arise dominantly from  covalent bonding or alternatively from an ionic one. For the former, exfoliation and twisting sheets of materials should work well, while for the latter it will be difficult to do the same. In addition, there is a large class whose bonding is intermediate between ionic and covalent. For those exfoliation and twisting techniques need to be tested experimentally.
}

\subsection*{Read-Out of the Ground State Properties}

Electrical transport has been the technique of choice to probe the ground state quantum properties of \moire structures. The first experiments in this direction were the measurements of quantum Hall effects in graphene on hBN, which revealed the influence of the \moire potential via the formation of Hofstadter butterfly patterns in the magnetotransport \cite{dean2013hofstadter}.

 A dramatic development was the discovery of collective quantum phases in twisted bilayer graphene near the so-called magic angle \cite{bistritzer2011}. After the original discovery of insulating  \cite{cao2018b} and superconducting phases \cite{cao2018a}, new  orbitally magnetized phases with a quantized Hall effect, indicating the development of topological phases that spontaneously break time reversal symmetry, have also been observed \cite{sharpe19,lu2019superconductors}. Similar collective and topological phases have been seen in the case of ABC graphene on hBN \cite{chen2019evidence}, in which case the \moire pattern between the ABC graphene and hBN provides additional flattening of the bandstructure which already displays a peak in the density of states at charge neutrality.

More recently, the methods applied to graphene have also been applied to  TMD-based heterostructures. The first experiments indicating the development of interaction-driven insulating phases have been performed in both TMD heterobilayers and homobilayers \cite{wang19,tang2020}.

\subsection*{Spectroscopic Read-Out}

\Moire structures present two great advantages for experiments in comparison to traditional solid state materials. The first is that the size of the "lattice" is expanded from the atomic scale to  several nanometers, allowing optical and near-field techniques to directly probe the local spectroscopic properties of the system, in close analogy with the quantum gas microscope invented in the case of cold atomic gases. The second advantage is in their intrinsic two-dimensional nature, which makes it possible to measure the properties of the entire sample with surface-sensitive spectroscopic probes such as scanning tunneling microscopy (STM) or angle-resolved photoemission. This does come with its disadvantages however, since many traditional spectroscopic probes applied to solid state systems (neutron spectroscopy for example) produce signals that are proportional to the volume of the sample being measured, and are thus not easily applicable to these materials.

 STM has been applied extensively to study the  bandstructure in twisted graphene structures \cite{Li2010,Kerelsky18,xie2019spectroscopic,jiang2019charge,Choi19}, to visualize internal edge states that exist at domain boundaries in small angle twisted bilayer and four-layer graphene systems, and to relate these states to the topological properties of the individual domains \cite{huang18b,kerelsky19}. Touching upon many-body correlation effects, prominent results obtained so far are a measurement of the insulating gap at half-filling, evidence that this insulating state does not break transnational symmetry but does break rotational symmetry near half-filling \cite{Kerelsky18,jiang2019charge,Choi19} and the discovery of an insulating phase at charge neutrality in four-layer rhombohedral graphene produced by twisting two bilayers \cite{kerelsky19}.

Furthermore, van der Waals heterostructures display nearly all optical phenomena found in solids, including plasmonic oscillations of free electrons characteristic of metals, light emission/lasing and excitons encountered in semiconductors, and intense phonon resonances typical of insulators \cite{Sunku1153,Low2017}. Therefore, optical spectroscopies allow one to reconstruct the role of \moire  superlattice potentials in the electronic structure of twisted multi-layers as well as in their electron and lattice dynamics. For example infrared spectroscopy on a graphene/boron nitride heterostructure shows that the \moire superlattice potential is dominated by a pseudospin-mixing component analogous to a spatially varying pseudomagnetic field \cite{Shi2014}, and photo-luminescence have uncovered the systematic evolution of light-emitting properties associated with interlayer excitons in TMD hetero-bilayers \cite{Seyler2019}.

Conventional diffraction-limited optics suffers from the spatial resolution restricted by the wave-length of light and therefore provides only area-averaged information on the electromagnetic response of vdW heterostructures. Modern nano-optical methods {\cite{Liu_2016,1910.07893}} allow one to overcome this limitation and enable optical inquiry at the length scale commensurate with \moire  periodicities. Additionally, nano-optical methods give access to spectroscopy and imaging of hybrid light-matter modes known as polaritons \cite{Sunku1153}. {Moir\'e structures also allow to set up a highly accurate tool enabling Moir\'e metrology of energy landscapes in 2d van der Waals structures \cite{2008.04835}}.

We conclude this subsection* by noting that functionalized AFM-tips enable multiple nano-scale contrasts in addition to nano-infrared studies including: nano-Raman, magnetic force microscopy, Kelvin probe and piezo force microscopy \cite{mcgilly19}. All these different methods provide complimentary insight carrying separate message about the studied phenomena \cite{2008.04835}. Co-located visualization of contrasts obtained with multiple functionalized scanning probes offers an opportunity for multi-messenger imaging of the interplay between electronic, magnetic and lattice effect at the lengths scale of moire domains \cite{McLeod2020}.

\section*{What Comes Next?}
\label{sec:next}
An important question is  which lattice structure might be realizable using Moir\'e physics in the future. Some promising future avenues of research are illustrated in Table \ref{tab:over} and in the outer circle of Fig.~\ref{fig1} -- while extensive, these lists are not exhaustive and we are confident that other low-energy model realizations will emerge as more heterostructure ``building blocks'' become experimentally viable.  We concentrate on a few highly sought after candidates next.

    \subsection*{Quantum Lego with Correlated or Topological Monolayers}
Beyond utilizing \textit{a priori} weakly correlated monolayers as overviewed above, a promising approach concerns constructing twisted heterostructures out of monolayers that by themselves already exhibit correlation-driven quantum phases. 
We present a short perspective of prototype 2D materials that should be studied in the future.
\begin{enumerate}
    \item In niobium diselenide (NbSe$_2$) superconducting behavior can be found even when exfoliated down to the monolayer (at temperatures of the order of 3 Kelvin) {\cite{xi2016ising}}. Two  such sheets might provide a solid-state-based inroad into the question of how the superconducting state emerges within a conventional BCS picture  and how this state crosses over to a strong coupling regime. 
    \item In the two-dimensional van der Waals material Chromium(III)iodide (CrI$_3$), quantum magnetic properties can be realized that depend on the number of layers and relative stacking order {\cite{soriano2020magnetic}}. Twisted heterostructures made of CrI$_3$ could permit accessing antiferromagnetically and ferromagnetically aligned domains in a flexible fashion using Moir\'e lattice engineering {\cite{hejazi2020noncollinear,wang2020stacking}}.
    \item {RuCl$_3$ realizes a Kitaev-Heisenberg magnet with stripe order {\cite{banerjee2016proximate}}. Layering on a substrate at a twist, or twisting layers of RuCl$_3$, can provide a method to tune the magnetic interactions and putatively tune the system towards a quantum spin liquid, as well as study the effects of doping a proximal spin liquid in a controlled manner.}
    \item TaSe$_2$ realizes a correlation-driven insulating state that is characterized by a star of David charge density wave configuration at low temperatures which yields a rather large unit cell [Fig. 1, outer circle] {\cite{chen2020strong}}. Again this correlation-driven phase is present even in the untwisted case. Twisting two of these structure atop each other raises the interesting questions of if and how superstructures made of these larger star of David charge density waves can be realized  and what are their manifestation in the electronic properties of the material.
    
    \item Beyond van der Waals materials the \moire approach to modify electronic structures is also at play in correlated oxides. These are epitaxial structures and therefore cannot be rotated easily. However a \moire pattern in films of the prototypical magnetoresistive oxide, La$_{0.67}$Sr$_{0.33}$MnO$_3$ can be epitaxially grown on LaAlO$_3$ substrates. The net effect is that both the electronic conductivity and ferromagnetism  are modulated by this \moire  engineering over mesoscopic scales \cite{McLeod2020}. This opens up yet another route in the combinatory space of chemical compositions to use in \moire systems.
    
    \item { 1T' WTe$_2$ monolayers realize a two-dimensional topological insulator due to a spin-orbit driven topological band inversion \cite{qian2014quantum}. While layering two WTe$_2$ monolayers would naively lead to a Z$_2$ trivial insulating state, the superlattice potential at finite twist angles can in principle induce a series of stacking-controlled topological transitions. A remarkable consequence of such a device would be the possibility to study the combination of non-trivial band topology and electronic correlations in Moir\'e bands with quenched dispersion, potentially giving rise to experimentally-realizable manifestations of fractional Chern insulators or fractional topological insulator phases.}
\end{enumerate}

    \subsection*{Highly Controllable Geometrically Frustrated Lattices: Kagome/Kitaev and spin liquids }
  One long-sought research goal regards the controlled realization of Kagome or Kitaev quantum magnets {\cite{balents2010spin}}. 
  In conventional candidate systems such as herbertsmithite, RuCl$_3$, iridates or metal-organic complexes, disorder-free realizations of the low-energy magnetic Hamiltonian or suppression of unwanted and longer-ranged spin exchange interactions are major challenges.
  A Moir\'e realization of the Kagome lattice or of strongly spin-orbit-coupled multi-orbital Moir\'e models in the strong-interaction limit would permit tuning magnetic interactions in a controlled manner, possibly allowing a realization of a quantum spin liquid state. { This would elevate \moire heterostructures as a platform to simulate an experimentally elusive phase of matter in a highly controllable setting, with a potentially worthwhile starting candidate being twisted layers of RuCl$_3$.}
  
  {Another direction recently put forward in the realization of an effective Kagome lattice, is twisted MoS$_2$, which was discussed above in the context of the asymmetric $p_x,p_y$ Hubbard model. Beyond this at very small angle, the next family of bands, which can be accessed by further doping the system, would effectively realize a multi-orbital generalization of a Kagome lattice \cite{xian20,2008.01735}. Future studies need to address whether such small angles and large degree of doping can be realized experimentally.   }
  
 \subsection*{   Proximity effects and spin orbit coupling}
 {One key advantage of two-dimensional systems is the possibility to induce effects via proximity to an engineered substrate. To avoid chemical modifications of the \moire system by the interface through strong chemical bonds, which would destroy the low-energy \moire bands of the original heterostructure, one can employ 2D heterostacking. } 
 Many interesting effects are expected in flat band systems, when superconducting or strongly spin-orbit coupled substrates are used to imprint some of their properties on the two-dimensional system under scrutiny.  For instance, large induced spin-orbit coupling {\cite{island2019spin}} might reveal novel topological superconducting states with applications to quantum computing. Similarly, proximitization with superconducting substrates could permit a controlled study of Majorana wires and lattices {\cite{liangfu08,lutchyn10,yuval10}}, e.g., via using a quasi one-dimensional Moir\'e material such as twisted GeSe. An important corrollary of exploiting a Moir\'e superstructure would be the delocalization of bound Majorana states on scales larger than the Moir\'e unit cell, promoting their length scale from Angstr\"oms to nanometers. Potential topological phenomena include the demonstration of a high-temperature quantum anomalous Hall effect and Majorana modes as well as antiferromagnetic topological insulators which open a route towards topological spintronics. 

\subsection*{From 2D to 3D}
Another intriguing future research direction concerns extending the idea of simulating quantum models from the one- or two-dimensional (discussed so far) to the three-dimensional realm using Moir\'e systems. To this end, a new exfoliation technique was recently demonstrated, to yield large-area atomically thin layers that can be stacked in any desired order and orientation to generate a whole new class of artificial materials, including stacking thick twisted materials \cite{liu2020disassembling}. We next discuss explicitly two examples, with obvious intriguing extensions possible.

\begin{enumerate}
\item Alternating twist:
When stacking many layers atop each other we want to define the twist angle of each layer to be measured with respect to the bottom one. Stacking two-dimensional materials at alternating twist then means that the  twist angle alternates between two values: zero and $\alpha$. If $\alpha$ is small, in-plane localized sites will emerge by the Moir\'e interference physics, and  these sites will lie directly atop each other in the out-of-plane direction.  The in-plane flat band states will become dispersive along the out-of-plane direction. Upon slight doping, the system thus becomes dominantly a one-dimensional metal with small residual in-plane coupling by the residual dispersion of the flat bands, similar to what has been discussed for bulk TaS$_2$ \cite{darancet14}. Such a one-dimensional metal is susceptible to instabilities (such as CDW, SDW or excitonic instabilities)  as a one-dimensional system is always perfectly nested. This will gap out the system along the out-of-plane direction and elevate the relevance of the small dispersion along the in-plane direction. A similar mechanism was recently observed to give rise to the 3D quantum Hall effect in bulk ZrTe$_5$ \cite{tang2019three} and similar fascinating surprises might be expected in the alternating twist configuration we propose here. 

\item Continuous twist:
Another intriguing idea is to stack layer by layer with constant twist angle $\alpha_0$ between adjacent layers, such that the above defined twist angle with respect to the bottom layer increases linearly $\alpha= \alpha_0 \cdot i {\;\rm{mod}\;}360^\circ$, where $i$ is the layer index. Calculations to characterize such a system are challenging as the unit cell w.r.t. the in-plane direction grows huge or a quasi-crystal structure with no unit cell emerges when more and more layers are added. Here the generalized Bloch band theory developed in \cite{wu2020three} might be useful. However, the general physics of emerging flat bands should transfer to the third direction, providing an inroad to realize 3D quantum models with tunable ratio of potential and kinetic energy scales as discussed above for the 2D case.

\end{enumerate}

\subsection*{Moir\'e Heterostructures out of Equilibrium}
So far we addressed equilibrium properties of twisted systems and we illustrated the rich phenomena that one can realize via engineering a variety of low-energy quantum many-body Hamiltonians. With increasingly sensitive pump-probe experiments available, another enticing possibility is to drive such systems out of equilibrium via short laser pulses or via embedding the material in a quantum cavity \cite{Ruggenthaler18,powerOfChirality,Giacomo19,sentef2018cavity,curtis18}. As the Moir\'e potential lowers the relevant kinetic energy scales to meV, these lie within reach of the typical energy scales of light-matter interactions, permitting outsized transient modifications of the electronic dynamics. Combining control of the twist angle with non-equilibrium perturbations  thus holds great promise to offer a new platform with opportunities to realize novel correlated non-equilibrium states of matter:

\begin{enumerate}
\item {Floquet engineering: in TBG, the sensitivity of correlated phases to meV changes of the dispersion of the Moire bands as a function of deviation from the magic angle suggests that rich effects can be achieved already via dressing the electronic bands by weak light fields. In the presence of an optical pump, a photon-mediated renormalization of the hopping matrix elements between Moir\'e unit cells can selectively modify the low-energy band structure and flatten its dispersion.}
\item {Cavity engineering: analogously, cavity engineering recently attracted great interest as a means to exploit strong light-matter coupling to change electronic phases. The comparatively low energy scales of Moir\'e bands and correlated phases in twisted bilayer graphene suggests that coupling to a cavity photon mode can induce an outsized effect on the interacting electronic state. One such application is in photo-induced and cavity-induced superconductivity. New predictions for quantum phenomena in these cavities include the enhancement of superconductivity  using the coupling between the vacuum fluctuations of the cavity and plasmonic modes of the environment \cite{sentef2018cavity,curtis18,schlawin2019,thomas19}, the study of strongly correlated Bose–Fermi mixtures of degenerate electrons and dipolar excitons \cite{shimazaki19,paravicinibagliani18,Li794,sun17,latini18}. }

\item Optically generated synthetic magnetic fields: The strong spin-orbit coupling in TMDs offers new possibilities to generate opto-magnetic quantum properties using dynamic fields. For instance, one can optically generate real magnetic fields from spin-valley polarization in TMD heterobilayers.  Under circularly polarized excitation at the bandgap of one of the TMD monolayers
in a heterobilayer, the initially large spin-orbit splitting can be further increased, leading to a high degree of spin-valley polarization of the hole. This, along with the lack of spin-valley polarization of the electron, will result in a net spin polarization, and thus, a real magnetic moment with the potential to induce novel non-equilibrium quantum phases.
\end{enumerate}

\subsection*{ Excitons at will}
Moir\'e platforms  also provide a basic infrastructure for achieving different interlayer exciton phases and excitonic lattices [see Fig. 1]. In particular, bilayers of TMDs were identified (even without twist) as intriguing candidates because the long lifetime of these charge-separated interlayer excitons \cite{Hong2014,Zhu2017,Jauregui870} should facilitate their condensation \cite{Wang2019}. Adding twist angle allows to realize Moir\'e exciton lattices { \cite{Seyler2019,Tran2019,Jin2019,Alexeev2019,shimazaki2020strongly}} which can feature topological exciton bands that support chiral excitonic edge states \cite{PhysRevLett.118.147401}.
The high degree of control permits study of the study the localized-delocalized crossover of excitons in Moir\'e lattices as well as a strain induced crossover from 2D to 1D exciton physics \cite{Bai19}. Minimizing the kinetic energy to increase  interactions is key to achieving robust condensation. E.g. twisted WSe$_2$ homobilayers have tunable flat electronic bands over a range of twist angles (see above), with alternatives including experimentally accessible heterobilayers as well as strain and pressure engineering. Two such band-engineered TMDs separated by a uniform spacer layer (hBN) {\cite{shimazaki2020strongly}} will likely provide the best avenue to achieve exciton condensation in zero applied field and to then tune the properties of the condensate by gating, field, pressure and strain. Future strategies to enhance the tendencies of excitons to condensate beyond the use of twist angle to achieve flat bands and thus more prominent interactions could also include the use of cavity structures to enhance light-matter interaction (see Fig. 1). Here, pump-probe nano-optical methods can both address and interrogate distinct regions in \moire heterostructures \cite{McLeod2020}. 

\subsection*{Skyrmions}

Skyrmions -- vortex-like magnetic configurations -- can arise as a spontaneously formed
hexagonal lattice in the ground state of chiral magnets, or as quasi-particle excitations on top of
a ferromagnetic ground state.  The energetic stability of skyrmions is a consequence of an anti-symmetric exchange (also called Dzyaloshinskii-
Moriya) interaction and they are interesting due to their non-trivial topological structure protecting them
against perturbations and rendering them highly attractive as potential memory devices. One candidate systems for which skyrmions have been predicted is monolayer CrI$_3$ {\cite{tong2018skyrmions}}. In the realm of twisted magnetic van der Waals heterostructures Moir\'e skyrmion lattices are favored due to the long-period Moir\'e pattern, which can be used to tune the periodicity and shape of the magnetization texture. In addition, an external magnetic field can be used to switch the vorticity and location of skyrmions.

\section*{Conclusion}
Here, we identified future directions of research into the blossoming fields of twisted van der Waals heterostructures with a particular emphasis on the high-degree of control achievable in these materials. Controlled engineering of quantum Hamiltonians as tunable low-energy theories of these structures is at the forefront of this fascinating research field. This idea might be viewed as a condensed matter realization of a quantum simulator in the future, which allows to directly access fundamental properties of quantum many-body systems, such as the delicate emergent physics of novel collective phases of matter, which have been experimentally elusive so far. In this sense twisted van der Waals heterostructures might be the long-sought after remedy to access the wider phase space, which should include all kinds of fundamentally and technologically relevant phases of matter.  { One of the major experimental challenges to be overcome next is to devise read-out techniques of the magnetic properties in these systems as well as to push the precision of angle resolved photoemission spectroscopy into the regime where it can resolve the physics within the small Brillouin zone of twisted heterostructures reliably. Both of these advances would provide cornerstones allowing to interrogate the low-energy physics of the system with regards to its interacting properties more completely.  }
Considering the tremendous combinatorial space of van der Waals materials to chose from, extensions to higher dimensions and the vibrant field of non-equilibrium control, including Floquet engineering and twist-tronics applied to these novel platforms, it is likely that only a small fraction of the potential phenomena achievable in twisted van der Waals heterostructures have been experimentally realized so far.  This newly emerging research field is likely to yield many more exciting development. 
 \\

\noindent \textbf{Acknowledgements:} 
This work is supported by the European Research Council (ERC-2015-AdG-694097) and Grupos Consolidados (IT1249-19). MC, AM, AG and AR are supported by the Flatiron Institute, a division of the Simons Foundation. We  acknowledge  funding 
by the Deutsche Forschungsgemeinschaft (DFG) under Germany's Excellence Strategy - Cluster  of  Excellence  Matter  and  Light  for Quantum Computing (ML4Q) EXC 2004/1 - 390534769, within the Priority Program SPP 2244 “2DMP” and Advanced Imaging of Matter (AIM) EXC 2056 - 390715994 and funding by the Deutsche Forschungsgemeinschaft (DFG) through RTG 1995 and RTG 2247. Support by the Max Planck Institute - New York City  Center for Non-Equilibrium Quantum Phenomena is acknowledged.
Work at Columbia is supported as part of Programmable Quantum Materials, an Energy Frontier Research Center funded by the U.S. Department of Energy (DOE), Office of Science, Basic Energy Sciences (BES), under award DE-SC0019443. DNB is the Vannevar Bush Faculty Fellow ONR-VB: N00014-19-1-2630.

\noindent \textbf{Competing interests:}
The authors declare no competing interests.


\end{document}